\documentclass[12pt]{iopart}

\usepackage{epsfig}
\usepackage{amssymb}

\newtheorem{theorem}{Theorem}
\newtheorem{lemma}[theorem]{Lemma}

\newcommand{\fa}{{f^{(a)}}}
\newcommand{\Fa}{{F^{(a)}}}
\newcommand{\fal}{{f_0^{(a)}}}
\newcommand{\far}{{f_1^{(a)}}}
\newcommand{\gal}{{g_0^{(a)}}}

\newcommand{\gar}{{g_1^{(a)}}}
\newcommand{\gaos}[1]{{g_{\underline{\omega}_{#1}}^{(a)}}}
\newcommand{\gazs}[1]{{g_{\underline{0}_{#1}}^{(a)}}}

\newcommand{\gao}[1]{{g_{\omega_{#1}}^{(a)}}}
\newcommand{\uo}[1]{\underline{\omega}_{#1}}
\newcommand{\lap}{\lambda^{(a)}_+ }
\newcommand{\lam}{\lambda^{(a)}_- }
\newcommand{\dx}{\mathrm{d}x\ }
\newcommand{\dy}{\mathrm{d}y\ }
\newcommand{\ds}{\mathrm{d}s\ }
\newcommand{\dt}{\mathrm{d}t\ }
\newcommand{\du}{\mathrm{d}u\ }
\newcommand{\dv}{\mathrm{d}v\ }
\newcommand{\qed}{\hfill$\blacksquare$}

\begin{document}

\paper[Time-reversible triangular maps of the square]
{Statistical properties of time-reversible triangular maps of the square}  
\author{Vasileios Basios, Thomas Gilbert\dag}
\address{Center for Nonlinear Phenomena and Complex Systems,
  Universit\'e Libre  de Bruxelles, CP~231, Campus Plaine, B-1050
  Brussels, Belgium}
\author{Gian Luigi Forti}
\address{Dipartimento di Matematica, Universit\`a degli Studi di Milano,
  via C Saldini, 50, I-20133 Milan, Italy}
\date{\today}

\begin{abstract}
Time reversal symmetric triangular maps of the unit square are introduced
with the property that the time evolution of one of their two variables is
determined by a piecewise expanding map of the unit interval. We study
their statistical properties and establish the conditions under which their
equilibrium measures have a product structure, \emph{i.e.} factorises in a
symmetric form. When these conditions are not verified, the equilibrium
measure does not have a product form and therefore provides additional 
information on the statistical properties of theses maps. This is the case
of anti-symmetric cusp maps, which have an intermittent fixed point and yet
have uniform invariant measures on the unit interval. We construct the
invariant density of the corresponding two-dimensional triangular map and
prove that it exhibits a singularity at the intermittent fixed point.  
\end{abstract}



\ead{\dag thomas.gilbert@ulb.ac.be}

\maketitle

It has long been realized that chaos is a ubiquitous property of non-linear
mechanical systems. The study of the dynamical properties of higher
dimensional systems such as those encountered in the framework of
statistical physics is however difficult from a theoretical stand-point,
and it is therefore rather naturally that the theory of chaotic dynamical
systems and their statistical properties was developed in the framework of
low-dimensional systems. The ergodic and dynamical properties of
one-dimensional piecewise expanding maps of the interval have been
extensively studied in this regard \cite{CE80} and, together
with Anosov diffeomorphisms, were central
to the derivation of some key results, in particular relating to the
Sina\"{\i}-Ruelle-Bowen theory of natural invariant measures. See references
\cite{B75,ER85}.    

A central problem in the study of statistical properties of
one-dimensional piecewise expanding maps is the identification of a natural
invariant measure. A standard procedure, which applies to maps with the
Markov property, is to establish a correspondence between the iterations of
such maps and shifts on semi-infinite sequences of spin variables. This way
the study of the invariant state of the map reduces to that of the 
corresponding lattice gas. One subsequently constructs 
a time translation invariant state by extending the semi-infinite spin
system to one which is infinite in both directions \cite{L81}. As far as the
one-dimensional (non-invertible) map is concerned, one might interpret this
procedure as restoring the symmetry under time reversal. In instances where
this procedure can be explicitly carried out at the level of the map, one
obtains a new time-reversal symmetric triangular map on the unit square, 
which reduces to the original one-dimensional map after projecting along
the appropriate dimension. 

To provide an example, it is well-known that the angle-doubling Bernoulli
map is equivalent to a coin-tossing game and that the equiprobability of all
sequences of heads and tails amounts to the invariance of the Lebesgue
measure on the interval. Extending the coin-tossing to memory-keeping
doubly-infinite sequences, one realizes that this construction amounts to
associating the baker map to the angle-doubling map, which is defined on
the square and is time-reversal invariant. See for instance
\cite{Be78}. Notice that the invariant density is uniform, both with the  
one-dimensional and two-dimensional maps. One might pedantically say that
the invariant density of the two-dimensional map is the product of that of
the one-dimensional map evaluated along the two dimensions of the
triangular map. Though this observation is trivial, it is indeed a case of
the triangular map having an invariant density with what will be referred
to as the product structure. This notion will play a central role in our
discussion. 

It is our purpose to show how this construction from one- to
two-dimensional maps can be generalized and what properties of the
invariant state can be inferred. Under specific symmetry 
assumptions on the map of the interval, we associate to it a
two-dimensional triangular map of the unit square, symmetric under
time-reversal. Considering the statistical properties of these maps, we
establish the conditions so that the invariant measure of the triangular
map has a smooth invariant density and identify the necessary and
sufficient conditions under which this invariant density has a product
structure in the sense defined above: namely it can be written as the
product of the density associated to the invariant measure of the
one-dimensional map, evaluated along both dimensions.

As we describe below, the case where the invariant measure of the
triangular map has a product form is arguably less interesting than when it
does not. The result indeed suggests that, unless the measure has a product
form, a complete statistical study of the expanding map of the interval
requires considering its time-reversible triangular extension to the unit
square. In other words there is more to learn about the statistical
properties of one-dimensional map by studying the statistics of the
associated two-dimensional map.

For the sake of illustration, we will consider in some detail a
one-parameter class of time-reversal symmetric triangular map with a
cusp. The interesting peculiarity of this class is that its limit upon
variation of the parameter becomes intermittent. Yet all the maps of the
class have equally uniform invariant densities on the unit interval
\cite{GS84}. This seemingly paradoxical property can be further explained
provided one considers the corresponding two dimensional triangular map. As
it turns out, the invariant measure develops a singularity as the parameter
value tends to the intermittent limit.  

The paper is organized as follows. The class of time-reversal symmetric
triangular maps we will consider is defined in section \ref{sec.model}. In
section \ref{sec.acm}, we prove that the invariant measure of these maps is
absolutely continuous. In section \ref{sec.dconj}, we consider the special
class of time-reversal symmetric triangular maps which are diffeomorphically
conjugated to maps that preserve the volume measure and show that this
condition is necessary and sufficient for the invariant measure to have the
product form. Section \ref{sec.cusp} is devoted to a one-dimensional
parameter class of time-reversal triangular maps which are generalizations
of the one-dimensional cusp maps and establish key properties of their
equilibrium states.

\section{Time-reversible triangular maps of the square \label{sec.model}}

Triangular or skew-product maps of the square are maps $F:[0,1]^2\mapsto
[0,1]^2$ of the 
form $F(x,y) = (f(x), g(x,y))$. A familiar such example is the baker map,
which expands the square horizontally by a factor of $2$ and
squeezes it vertically by $1/2$ so as to preserve areas, and subsequently
folds the two horizontal halves on top of one another, thus recovering the
unit square. Specifically, it is defined according to  
\begin{equation}
B\ :\ (x,y)\mapsto \left\{
\begin{array}{l@{\quad}l}
(2x,\frac{y}{2}),&0\leq x < 1/2,\\
(2x-1,\frac{y+1}{2}),&1/2\leq x < 1.
\end{array}
\right.
\end{equation}
An important property of the baker map is that it is time-reversal
symmetric, {\em i.~e.} there exists an involution of the unit square,
$T:[0,1]^2\mapsto [0,1]^2$ such that 
\begin{equation}
T\circ B \circ T = B^{-1}.
\end{equation}
The baker map has actually two such symmetries, which map the square along
its respective diagonals, $T_1(x,y) = (y,x)$ and $T_2(x,y) = (1-y,1-x)$.

It is the purpose of this paper to establish the statistical properties of
triangular maps of the square which are time reversal symmetric and can be
obtained from the baker map whether through conjugation or continuous
deformation. 

More specifically, we consider maps of the form
\begin{equation}
F\ :\ (x,y)\mapsto
\left\{
\begin{array}{l@{\quad}l} 
\big(f_0(x),g_0(y)\big),&0\leq x < 1/2,\\
\big(f_1(x),g_1(y)\big),&1/2\leq x < 1,
\end{array}
\right.
\label{Fmap}
\end{equation}
where, on the one hand, $f_0(x)$ is twice differentiable, strictly expanding,
{\em i.~e.} $f_0'(x) \ge \alpha > 1$, with $f_0(0) = 0$, $f_0(1/2) = 1$,
and 
\begin{equation}
f_1(x) = 1 - f_0(1 - x).
\end{equation}
The $y$ component of $F$, on the other hand,
is defined through the inverse maps 
\begin{equation}
  \eqalign{
    g_0(x) = f_0^{-1}(x),\\
    g_1(x) = f_1^{-1}(x).
  } 
\end{equation}
By construction, maps (\ref{Fmap}) have the hyperbolic
properties and time-reversal symmetries of the baker map. 

\section{\label{sec.acm}Absolutely continuous measure}

Asymptotic statistical properties of maps (\ref{Fmap}) are determined by
absolutely continuous measures whose densities $\rho(x,y)$ are invariant
under the Perron Frobenius operator
\begin{equation}
  \rho(x, y) = g_\omega'(x) f_\omega'(y) 
  \rho\big(g_\omega(x), f_\omega(y) \big),
  \label{Frhoinv}
\end{equation} 
where $\omega = 0$ if $0\leq y <1/2$ and $\omega = 1$ if $1/2\leq y <1$.
This is the content of our first 
\begin{theorem}
  \label{thm.smoothrho}
  Let $F$ be a triangular map of the square of the form (\ref{Fmap}), as
  specified above. The natural invariant measure of $F$ is unique and
  absolutely continuous with respect to the volume measure, with density
  $\rho(x,y)$ whose marginals are equal,
  \begin{equation}
    \int_0^1\dy \rho(x,y) =   \int_0^1\dy \rho(y, x) \equiv \zeta(x),
    \label{marginalrho}
  \end{equation}
  where $\zeta(x)$ is the invariant density of the the one-dimensional map
  of the interval
  \begin{equation}
    f(x) =
    \left\{
      \begin{array}{l@{\quad}l}
        f_0(x),& 0\leq x < 1/2, \\
        f_1(x),& 1/2\leq x < 1.
      \end{array}
    \right.
    \label{1dmap}
  \end{equation}
\end{theorem}

To prove this result, we first notice that $f$ as defined by (\ref{1dmap})
is an expanding piecewise $C^2$ map of the interval, so that, by the theorem
of Lasota and Yorke \cite{LY73}, there is a unique $L^1$ integrable 
fixed point of the Perron-Frobenius operator, which we denote $\zeta(x)$,
\begin{equation}
  \zeta(x) = g_0'(x) \zeta(g_0(x)) + g_1'(x) \zeta(g_1(x)).
  \label{PFopf}
\end{equation}
This $\zeta(x)$ is the density associated to the natural invariant measure
of $f$. 

By extension, and since $F$ is triangular with $f$ its projection along the
unstable direction, $F$ has a unique Sinai-Ruelle-Bowen measure whose
conditional measure along the unstable direction has density $\zeta$, see
\cite{ER85} for a general discussion. Thus let $\rho$ denote the density
associated to the SRB measure of $F$. The relation $\zeta(x) = \int_0^1 \dy
\rho(x,y)$ ensues.  

The second equality of equation (\ref{marginalrho}), namely
$\zeta(y) = \int_0^1 \dx \rho(x,y)$, is less immediate and can be derived
starting with (\ref{Frhoinv}) evaluated at $(x, g_\omega(y))$. Integrating
over $x$, we obtain
\begin{equation}
  \eqalign{
    \int_0^1 \dx \rho(x,g_0(y)) = f_0'(g_0(y)) \int_0^{1/2} \dx \rho(x,y),\\
    \int_0^1 \dx \rho(x,g_1(y)) = f_1'(g_1(y)) \int_{1/2}^1 \dx \rho(x,y).
  }
  \label{xintrho}
\end{equation}
Using $f_\omega'(g_\omega(y)) = 1/g_\omega'(y)$, we can combine these two
equations and infer the relation
\begin{equation}
  \int_0^1 \dx \rho(x,y) = g_0'(y) \int_0^1\dx \rho(x,g_0(y)) 
  + g_1'(y) \int_0^1\dx \rho(x,g_1(y)),
\end{equation}
which completes the proof of equation (\ref{marginalrho}).

Its marginals being identical, the invariant density therefore has the
symmetries of $F$,  
\begin{equation}
  \rho(x,y) = \rho(y,x) = \rho(1- y, 1 - x) =  \rho(1- x, 1 - y). 
  \label{symrho}
\end{equation}

To finish the proof that $\rho(x,y)$ is absolutely continuous, it is
sufficient to show that the phase-space contraction rate of $F$, or sum of
the Lyapunov exponents, vanishes in average. Let $\lambda_+$ denote the
positive 
Lyapunov exponent of $F$, equal to that of $f$ above~:
\begin{equation}
  \lambda_+ = \int_0^{1/2} \dx \zeta(x) \log f_0'(x) 
  + \int_{1/2}^1 \dx \zeta(x) \log f_1'(x) 
  \label{lambdap}
\end{equation}

The negative Lyapunov exponent is 
\begin{equation}
  \lo \lambda_- = \int_0^{1/2}\dx \int_0^1 \dy \rho(x, y) \log g_0'(y) +
  \int_{1/2}^1\dx \int_0^1 \dy \rho(x, y) \log g_1'(y),
  \label{lambdam1}
\end{equation}
From equation (\ref{xintrho}) above, 
$\int_0^{1/2}\dx \rho(x, y) = g_0'(y) \zeta(g_0(y))$ and
$\int_{1/2}^1\dx \rho(x, y) = g_1'(y) \zeta(g_1(y))$.
Therefore
\begin{eqnarray}
\lo \lambda_- = \int_0^1 \dx \bigg[g_0'(x) \zeta(g_0(x)) \log g_0'(x) +
g_1'(x) \zeta(g_1(x)) \log g_1'(x)\bigg], \nonumber\\
= \int_0^{1/2} \dx \zeta(x) \log g_0'(f_0(x)) +
\int_{1/2}^1 \dx \zeta(x) \log g_1'(f_1(x)), \nonumber\\
= - \int_0^{1/2} \dx \zeta(x) \log f_0'(x) -
\int_{1/2}^1 \dx \zeta(x) \log f_1'(x), \nonumber\\
= - \lambda_+,
\label{lambdam2}
\end{eqnarray}
where, in the third line, we used the identity $g_\omega'(f_\omega(x)) = 
1/f_\omega'(x)$.

\qed

Triangular maps constructed upon one-dimensional differentiable expanding
maps of the circle which are symmetric about $1/2$ therefore have
absolutely continuous invariant measures. 

Notice that, upon inspection of equation (\ref{Frhoinv}), one might be led
to believe that $\rho(x,y)$ has the product structure,  
\begin{equation}
\rho(x,y) = \zeta(x) \zeta(y).
\label{rho2dprod}
\end{equation}
Indeed, since $f$ and $g$ are the inverses of one another, one
may substitute $g_\omega(z)$ ($\omega = 0,1$, $0\leq z < 1$) for $y$ in
equation (\ref{Frhoinv}) and write 
\begin{equation}
g_\omega'(z) \rho\big(x, g_\omega(z)\big) = 
g_\omega'(x) \rho\big(g_\omega(x), z \big).
\label{Frhoinvb}
\end{equation}
This equation is symmetric between $x$ and $z$, except for the value of
$\omega$, which is determined according to that of $y$: 
\begin{equation}
  \eqalign{
    \omega = 0,\quad 0\leq y<1/2,\\
    \omega = 1,\quad 1/2\leq y<1.\\
  }
\end{equation}
Therefore the product form (\ref{rho2dprod}) is in general invalid,
unless equation (\ref{Frhoinvb}) is independent of $z$, which requires the
identity
\begin{equation}
g_0'(z) \zeta\big(g_0(z)\big) = g_1'(z) \zeta\big(g_1(z)\big).
\label{rhoprodsolcond}
\end{equation}

As we will demonstrate shortly it is easy to find maps $F$ of the form
(\ref{Fmap}) which do not verify this property and consequently do not 
have a product measure. 

The question which we
address in the next section is to determine under which conditions the
density $\rho(x,y)$ has the product form (\ref{rho2dprod}).

\section{\label{sec.dconj}Diffeomorphic conjugations}

An example of a non-trivial two-dimensional map for which equation
(\ref{rhoprodsolcond}) holds is that of the anti-symmetric logistic
map. Let $f_0(x) = 4x(1-x)$ and $f_1$ defined as in equation
(\ref{f1}). The inverses are $g_0(x) = 1/2(1-\sqrt{1-x})$ and $g_1(x) =
1/2(1+\sqrt{x})$ respectively.  The invariant density of the
one-dimensional map (\ref{1dmap}) is $\zeta(x) = 1/[\pi \sqrt{x(1-x)}]$ 
and one easily checks that $\rho(x,y) = \zeta(x)\zeta(y)$ verifies equation
(\ref{Frhoinv}).

This property can be understood as a consequence of the diffeomorphic
conjugation of the map above to the baker map. As is well-known
\cite{LMckey94}, the one-dimensional logistic map above can be obtained
from the angle doubling map by a conjugation, 
\begin{equation}
  f(x) = \phi^{-1}(2\phi(x)\ \mathrm{mod}\ 1),
  \label{fconj2x}
\end{equation} 
where $\phi(x) = 1-2/\pi \arccos\sqrt{x}$, with inverse $\phi^{-1}(x) =
\cos^2[\pi (1-x)/2]$, and such that $\phi'(x) = \zeta(x)$.

We now set on to establish the generality of this result. Thus consider $F$
as defined in (\ref{Fmap}) and let $\phi$ be a diffeomorphism of the unit
interval with positive derivative. Set $\phi'(x) \equiv \sigma(x)$;
$\sigma$ is a positive density. Consider now the function
\begin{equation}
\widehat{F}(x,y) = 
\left\{
\begin{array}{l@{\quad}l}
\bigg(\phi\circ f_0 \circ \phi^{-1} (x), 
\phi\circ g_0 \circ \phi^{-1} (x)\bigg),&
0\leq x < 1/2,\\
\bigg(\phi\circ f_1 \circ \phi^{-1} (x), 
\phi\circ g_1 \circ \phi^{-1} (x)\bigg),&
1/2 \leq x < 1.
\end{array}
\right.
\label{Fhat}
\end{equation}
We further set $\widehat{f}_\omega \equiv \phi\circ f_\omega \circ
\phi^{-1}$ and $\widehat{g}_\omega \equiv \phi\circ g_\omega \circ
\phi^{-1}$, $\omega = 0,1$.

\begin{lemma}
  Let $\mathcal{P}_F$ and $\mathcal{P}_{\widehat{F}}$ denote the Perron
  Frobenius operators corresponding to $F$ and $\widehat{F}$
  respectively. Then
  \begin{equation}
    \mathcal{P}_F \eta(x,y) = \mathcal{P}_\phi^{-1}
    \bigg[\mathcal{P}_{\widehat{F}}\bigg( \mathcal{P}_\phi \eta \bigg)(x,y)
    \bigg],
    \label{PFPphi}
  \end{equation}
  where 
  \begin{equation}
    \mathcal{P}_\phi \eta (x,y) = \frac{\eta\bigg(\phi^{-1}(x),
      \phi^{-1}(y) \bigg)}{\sigma\bigg(\phi^{-1}(x)\bigg)
      \sigma\bigg(\phi^{-1}(y)\bigg)} 
    \label{Pphi}
  \end{equation}
\end{lemma}

We have 
\begin{equation}
  \int_0^x \ds \int_0^y \dt \mathcal{P}_F \eta(s,t) = 
  \int \int_{F^{-1}\big([0,x]\times[0,y]\big)} \ds \dt \eta(s, t),
\end{equation}
where
\begin{equation}
  F^{-1}\big([0,x]\times[0,y]\big) = 
  \left\{
    \begin{array}{l@{\quad}l}
      \big[0, g_0(x)\big] \times \big[0,f_0(y)\big], & 0 \leq y < 1/2,\\
      \big[0, g_0(x)\big] \times \big[0,1\big] \\
      \hspace{.5cm} \cup \big[1/2, g_1(x)\big]
      \times\big[0, f_1(y)\big], & 1/2 \leq y < 1.
    \end{array}
  \right.
\end{equation}
Assuming $0\leq y < 1/2$, we can write
\begin{eqnarray}
  \int_0^x \ds \int_0^y \dt \mathcal{P}_F \eta(s, t)
  &=& \int_0^{g_0(x)} \ds \int_0^{f_0(y)} \dt \eta(s, t),\nonumber\\
  &=& \int_0^{\phi^{-1}\circ\widehat{g}_0\circ\phi(x)} \ds 
  \int_0^{\phi^{-1}\circ\widehat{f}_0\circ\phi(y)} \dt \eta(s, t).
\end{eqnarray}
Setting $s = \phi^{-1}(u)$ and $t = \phi^{-1}(v)$, the Jacobian of the
transformation is $J = \phi'(s)\phi'(t) = \sigma(s) \sigma(t)$, so that the
last integral can be rewritten
\begin{eqnarray}
\fl \int_0^{\widehat{g}_0\circ\phi(x)} \du
\int_0^{\widehat{f}_0\circ\phi(y)} \dv
\frac{\eta\big(\phi^{-1}(u), \phi^{-1}(v)\big)}
{\sigma\big(\phi^{-1}(u)\big) \sigma\big(\phi^{-1}(v)\big)}
= \int_0^{\widehat{g}_0\circ\phi(x)} \du
\int_0^{\widehat{f}_0\circ\phi(y)} \dv 
\mathcal{P}_\phi\eta(u,v),\nonumber\\
= \int_0^{\phi(x)} \du \int_0^{\phi(y)} \dv
\mathcal{P}_{\widehat{F}}\big(\mathcal{P}_\phi \eta\big)(u, v),\nonumber\\
=\int_0^{x} \ds \int_0^{y} \dt
\mathcal{P}_\phi^{-1}\big[
\mathcal{P}_{\widehat{F}}\big(\mathcal{P}_\phi \eta\big)(s,t)\big].
\end{eqnarray}

This result holds for every $\eta \in L^1$. We therefore have derived the
relation 
\begin{equation}
  \mathcal{P}_F \eta(x, y) = 
  \mathcal{P}_\phi^{-1}\big[
  \mathcal{P}_{\widehat{F}}\big(\mathcal{P}_\phi \eta\big)(x, y)\big].
\end{equation}
The same relation holds for $1/2 \leq y < 1$.
\qed

\begin{theorem}
  The invariant density $\rho(x,y)$ of $F$ has the product form 
  \begin{equation}
    \rho(x,y) = \sigma(x)\sigma(y) = \phi'(x)\phi'(y).
    \label{prodrho}
  \end{equation}
  if and only if $\widehat{F}$, which is obtained from the conjugation of
  $F$ and $\phi$, is measure-preserving, {\em i.~e.} $\widehat{F}$ has
  uniform invariant density, $\mathcal{P}_{\widehat{F}} 
  1 = 1$.
\end{theorem}
We have
\begin{equation}
  \mathcal{P}_\phi \big(\sigma(x)\sigma(y)\big) = 
  \frac{\sigma\big(\phi^{-1}(x)\big)\sigma\big(\phi^{-1}(y)\big)}
  {\sigma\big(\phi^{-1}(x)\big)\sigma\big(\phi^{-1}(y)\big)} = 1.
\end{equation}
Thus, by the previous lemma,
\begin{eqnarray}
  \mathcal{P}_F\big(\sigma(x)\sigma(y)\big)
  &=& \mathcal{P}_\phi^{-1}\big[\mathcal{P}_{\widehat{F}}
  \big(\mathcal{P}_\phi\sigma(x)\sigma(y)\big)\big],\nonumber\\
  &=& \mathcal{P}_\phi^{-1}\big[\mathcal{P}_{\widehat{F}}
  1\big],\nonumber\\
  &=& \mathcal{P}_\phi^{-1} 1,\nonumber\\
  &=& \sigma(x)\sigma(y).
\end{eqnarray}

The converse is true. Assume that $F$ has invariant density $\rho(x,y) =
\zeta(x)\zeta(y)$ and consider the homeomorphism 
\begin{equation}
  \phi(x) = \int_0^x \ds \zeta(s).
\end{equation}
Then
\begin{equation}
  \mathcal{P}_\phi \rho(x,y) = \frac{\zeta\big(\phi^{-1}(x)\big)
    \zeta\big(\phi^{-1}(y)\big)}{\zeta\big(\phi^{-1}(x)\big)
    \zeta\big(\phi^{-1}(y)\big)} = 1.
\end{equation}
Let $\widehat{F}$ be the function constructed from $F$ and the
homeomorphism $\phi$. We then have
\begin{eqnarray}
  \zeta(x)\zeta(y) &=& \mathcal{P}_F\big( \zeta(x)\zeta(y) \big),
  \nonumber\\ 
  &=& \mathcal{P}_\phi^{-1}\big[\mathcal{P}_{\widehat{F}}1\big].
\end{eqnarray}
Assume that $\mathcal{P}_{\widehat{F}} 1 \equiv d(x,y) \neq 1$. Then
\begin{equation}
  \zeta(x)\zeta(y) = d\big(\phi(x), \phi(y)\big) \zeta(x)\zeta(y),
\end{equation}
a contradiction.
\qed

The invariant measures of maps that are conjugated to piecewise linear maps
such as the baker map  have therefore a product structure. And, conversely,
a map whose invariant measure has a product structure is conjugated to a
piecewise linear map by a diffeomorphism whose derivative is equal to the
marginals of its probability density.

In the next section, we turn to a one-parameter family of maps (\ref{Fmap})
that does not have this property. They are obtained by continuous
deformation of baker maps and, as the parameter is varied, have a singular
limit which displays intermittency. As it will turn out, the marginals are
uniform for all values of the parameter, yet the measure develops a
singularity as one approaches the intermittent regime.
  
\section{\label{sec.cusp}Cusp maps}

\begin{figure}[htb]
  \centerline{
    \includegraphics[width=.5\textwidth]{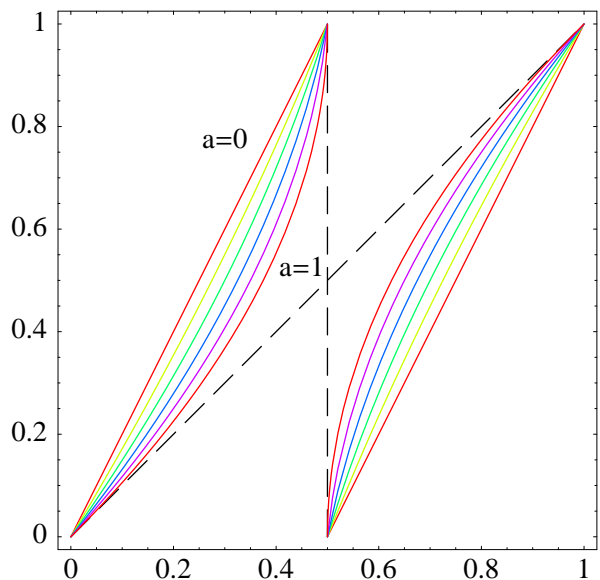}
  }
  \caption{Anti-symmetric cusps maps (\ref{1dcusp}), here shown for
    $a=0,1/5,\dots, 1$.}
  \label{fig.1dmaps}
\end{figure}

We consider a two-dimensional extension of a class of anti-symmetric cusp
maps, whose symmetric version was previously introduced in \cite{GS84}. Let
$0<a\le 1$. We define
\begin{equation}
\fal(x) = \frac{a+1}{2a}\left[
1 - \sqrt{1 - \frac{8 a x}{(a + 1)^2}}\right],
\label{f0}
\end{equation}
for $0\leq x\leq 1/2$, and
\begin{equation}
\far(x) = 1 - \fal(1-x),
\label{f1}
\end{equation}
for $1/2\leq x\leq 1$. The class of anti-symmetric cusp maps of the
interval is defined by  
\begin{equation}
\fa:\ x\mapsto
\left\{
\begin{array}{l@{\quad}l}
\fal(x),&0\leq x \leq 1/2,\\
\far(x),&1/2 < x \leq 1.
\end{array} 
\right.
\label{1dcusp}
\end{equation}
Figure \ref{fig.1dmaps} displays several of these functions as the
parameter $a$ is varied between 0 and 1.

The inverses of $\fal$ and $\far$ are 
\begin{equation}
  \eqalign{
    \gal(x) \equiv \fal^{-1} = \frac{1 + a}{2} x - \frac{a}{2} x^2,\\
    \gar(x) \equiv \far^{-1} = \frac{1}{2} + \frac{1 - a}{2} x + \frac{a}{2}
    x^2. 
  }
  \label{ga}
\end{equation}

An immediate property of the Perron-Frobenius operators attached to the
maps (\ref{1dcusp}) is that they preserve the Lebesgue measure and therefore
have uniform density, irrespective of the value of $a$. This follows from 
equation (\ref{PFopf}) and the identity
\begin{equation}
\gal'(x) + \gar'(x) 
= \frac{a + 1}{2} - a x + a x + \frac{1 - a}{2}
= 1,
\end{equation}
where the prime indicates the derivative with respect to the argument. Thus
the computation of the positive Lyapunov exponent is straightforward and yields
\begin{equation}
\lap = \log 2 + \frac{1}{2} + \frac{(1 - a)^2}{4 a}\log (1 - a)
- \frac{(1 + a)^2}{4 a}\log (1 + a).
\label{lyapp}
\end{equation}
In particular $\lambda^{(0)}_+ = \log 2$ and $\lambda^{(1)}_+ = 1/2$.

The specificity of the maps $\fa$ thus defined is that they can be viewed,
as one tunes the value of the parameter $a$ from $0$ to $1$, as continuous
deformations of the angle-doubling map, $x \mapsto  2 x$ if $0\leq x <
1/2$, or $2 x -1$ if $1/2\leq x < 1$, to the intermittent anti-symmetric
cusp map, $x \mapsto 1 - \sqrt{1 - 2 x}$ or $\sqrt{2 x - 1}$, depending on
whether $x < 1/2$ or $x \ge 1/2$. The latter case is weakly intermittent in
the sense that the slope of $f^{(1)}$ at $x=0$ is unity, and yet 
the invariant density is constant and therefore shows no sign of the
singularity that underlies the intermittency of its statistical
observables. Nevertheless this regime is  
characterized by the power law decay of correlation functions, which stems
from the existence of an accumulation of the eigenvalue spectrum of the
Frobenius-Perron operator towards the eigenvalue $1$, which corresponds to
the stationary state. See refs. \cite{He84,GH85,klb96,lbk96,bkl97,McK08}
for the treatment of the symmetric case.  

In order to display the effect of the arising intermittency as $a\to1$ on
the statistical properties of trajectories driven by the maps
(\ref{1dcusp}) and the singularity of the invariant measure in the
intermittent regime ($a=1$), one needs to consider the two-dimensional
extension of $\fa$  to time-reversible triangular maps, defined in
accordance to equation (\ref{Fmap})~:
\begin{equation}
\Fa:\ (x,\, y)\mapsto
\left\{
\begin{array}{l@{\quad}l}
\bigg(\fal(x),\ \gal(y)\bigg),& 
0\leq x \leq 1/2,\\
\bigg(\far(x),\ \gar(y)\bigg),& 
1/2 < x \leq 1.
\end{array} 
\right.
\label{2dcusp}
\end{equation}

By construction, $\Fa$, except at $a=1$, has the properties studied in
section \ref{sec.acm} and, in particular, verifies theorem
\ref{thm.smoothrho}, asserting that the invariant density is smooth, with
both marginals trivial. In particular $\lam = - \lap$.

From equation (\ref{Frhoinv}), the invariant density verifies the
functional equation 
\begin{eqnarray}
  \fl\rho_a(x, y) =
  \left\{
    \begin{array}{l@{\quad}l}
      \frac{1 + a - 2 a x}{\sqrt{(1 + a)^2 - 8 a y}}
      \rho_a\left(\frac{1 + a}{2}x - \frac{a}{2}x^2, 
        \frac{1 + a - \sqrt{(1 + a)^2 - 8 a y}}{2a}\right),&
      0\leq y \leq 1/2,\\  
      \frac{1 - a + 2 a x}{\sqrt{(1 + a)^2 - 8 a (1 - y)}}
      \rho_a\left(\frac{1}{2} + \frac{1 - a}{2}x + \frac{a}{2}x^2, 
        \frac{\sqrt{(1 + a)^2 - 8 a (1 - y)} - 1 + a}{2a}\right),&
      1/2\leq y \leq 1.  
    \end{array} 
  \right.
  \nonumber\\
  \label{rho2d}
\end{eqnarray}

\begin{figure}[htb]
  \centerline{
    \includegraphics[width=.45\textwidth]{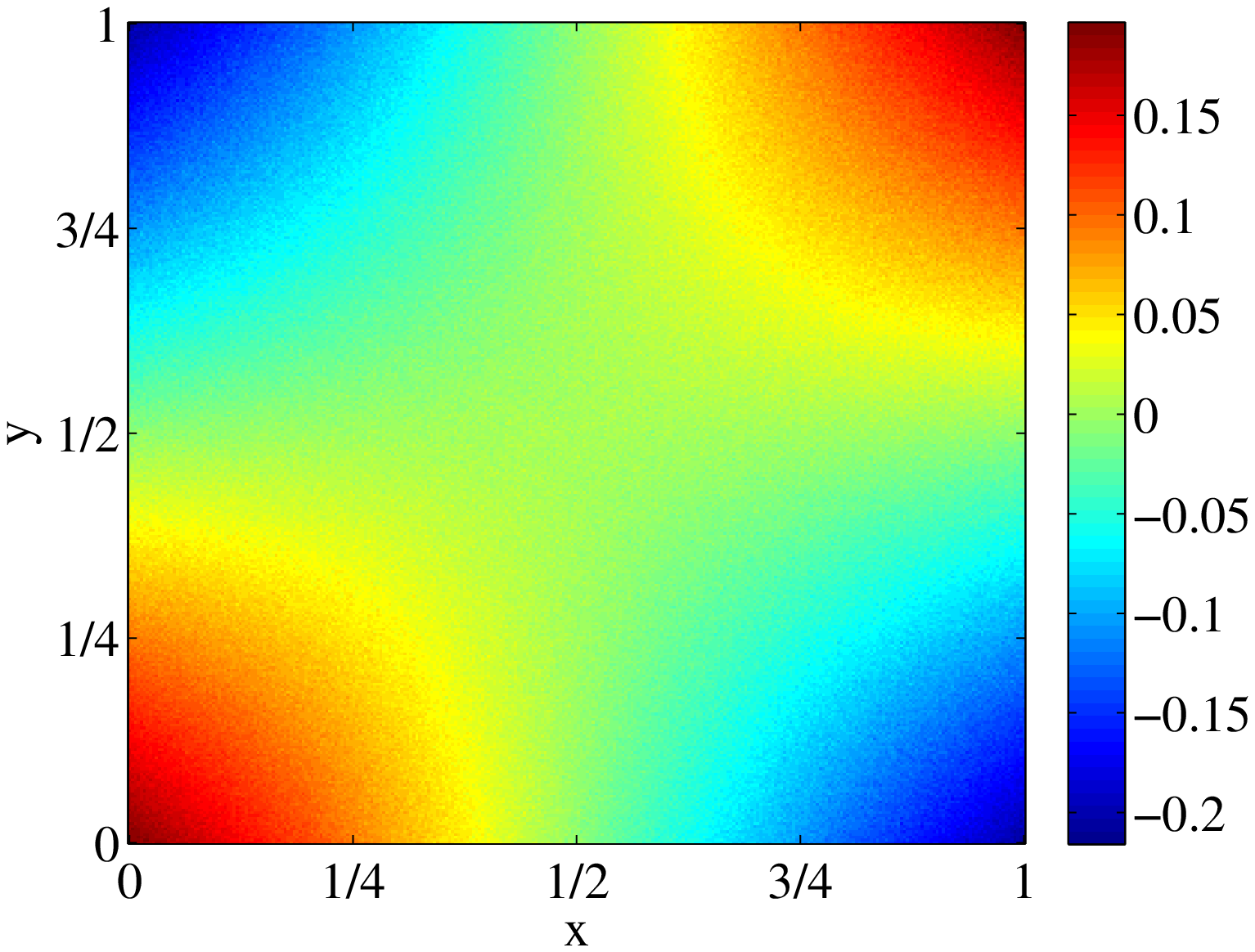}
    \hspace{.5cm}
    \includegraphics[width=.45\textwidth]{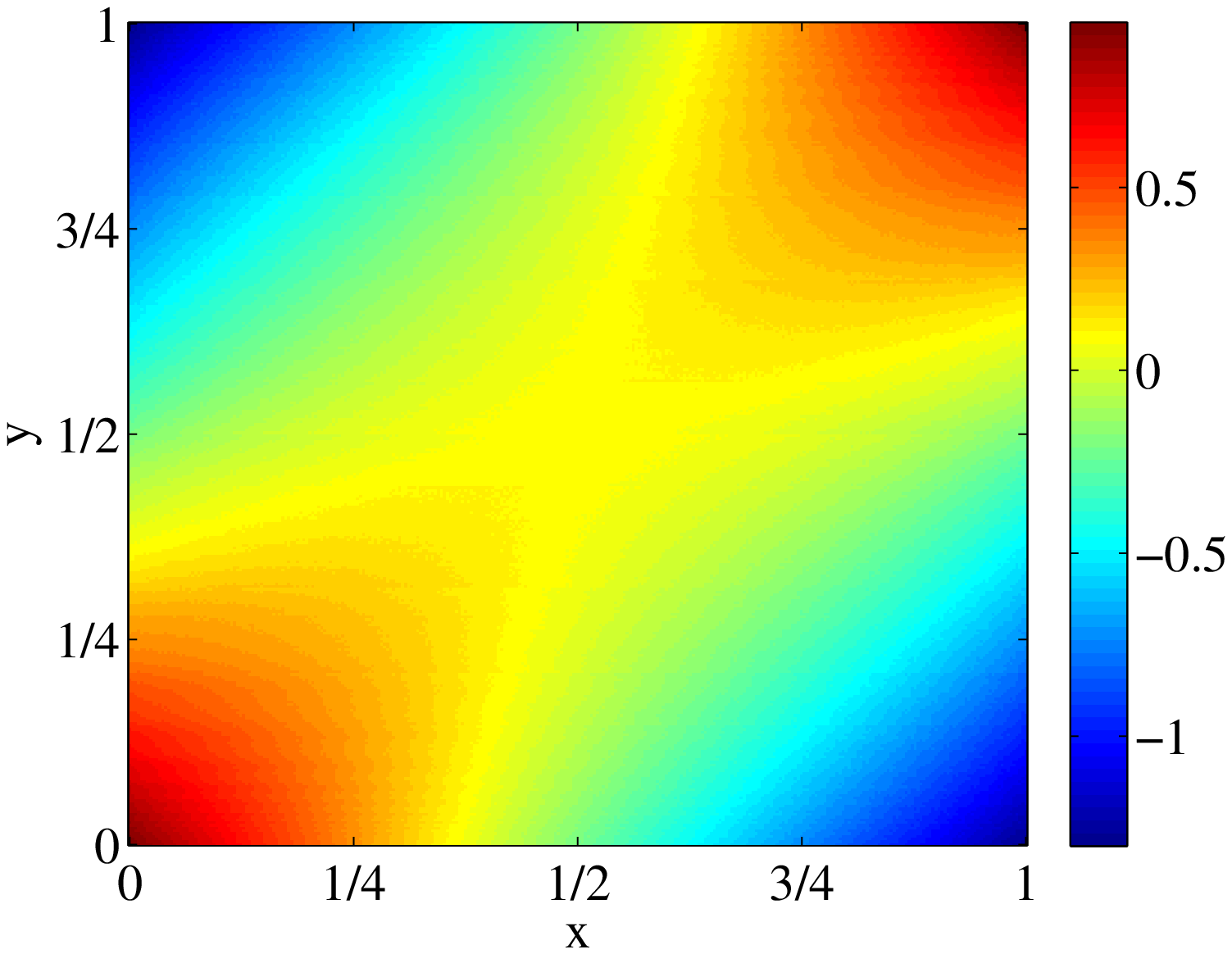}
  }
  \centerline{
    \includegraphics[width=.45\textwidth]{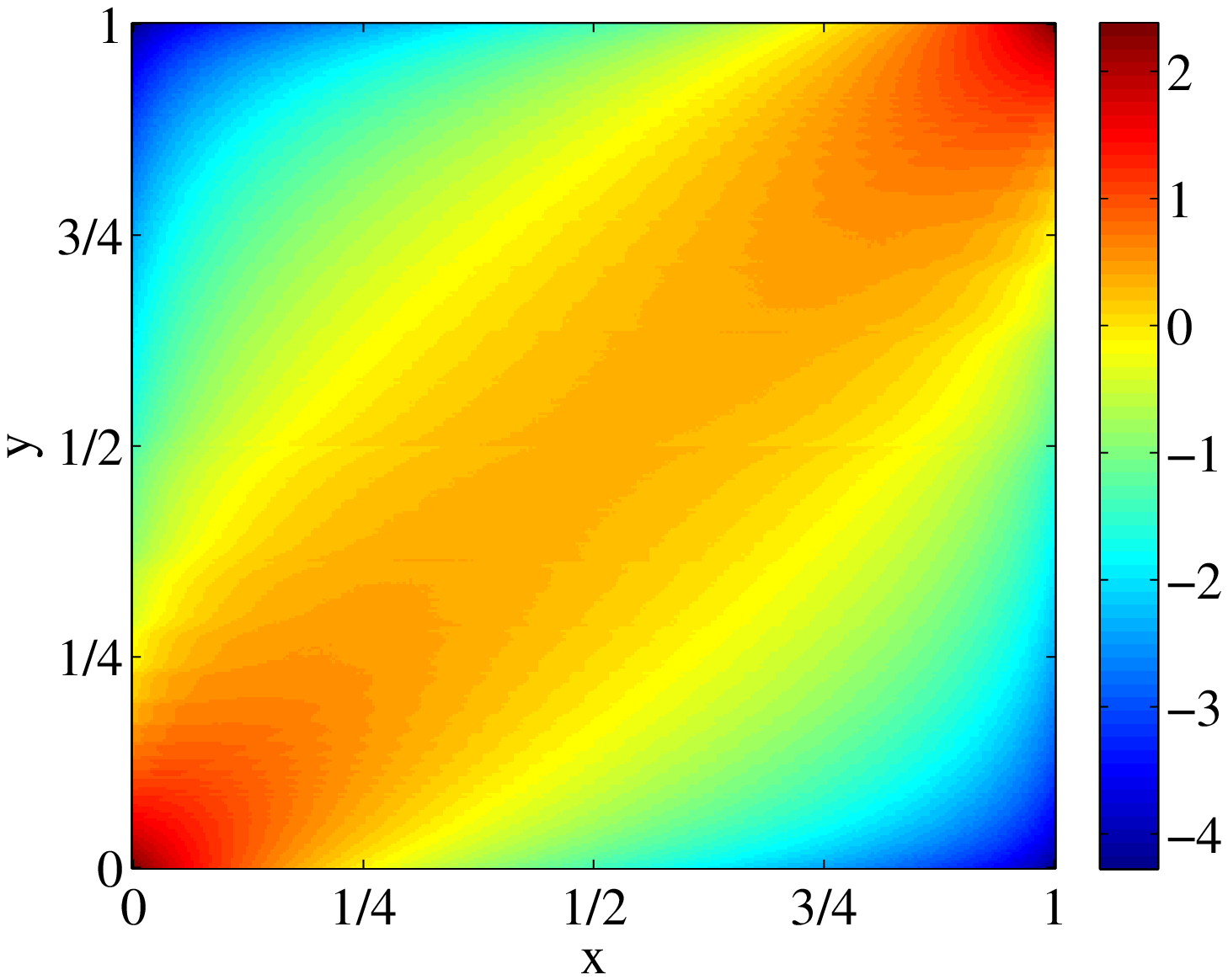}
    \hspace{.5cm}
    \includegraphics[width=.45\textwidth]{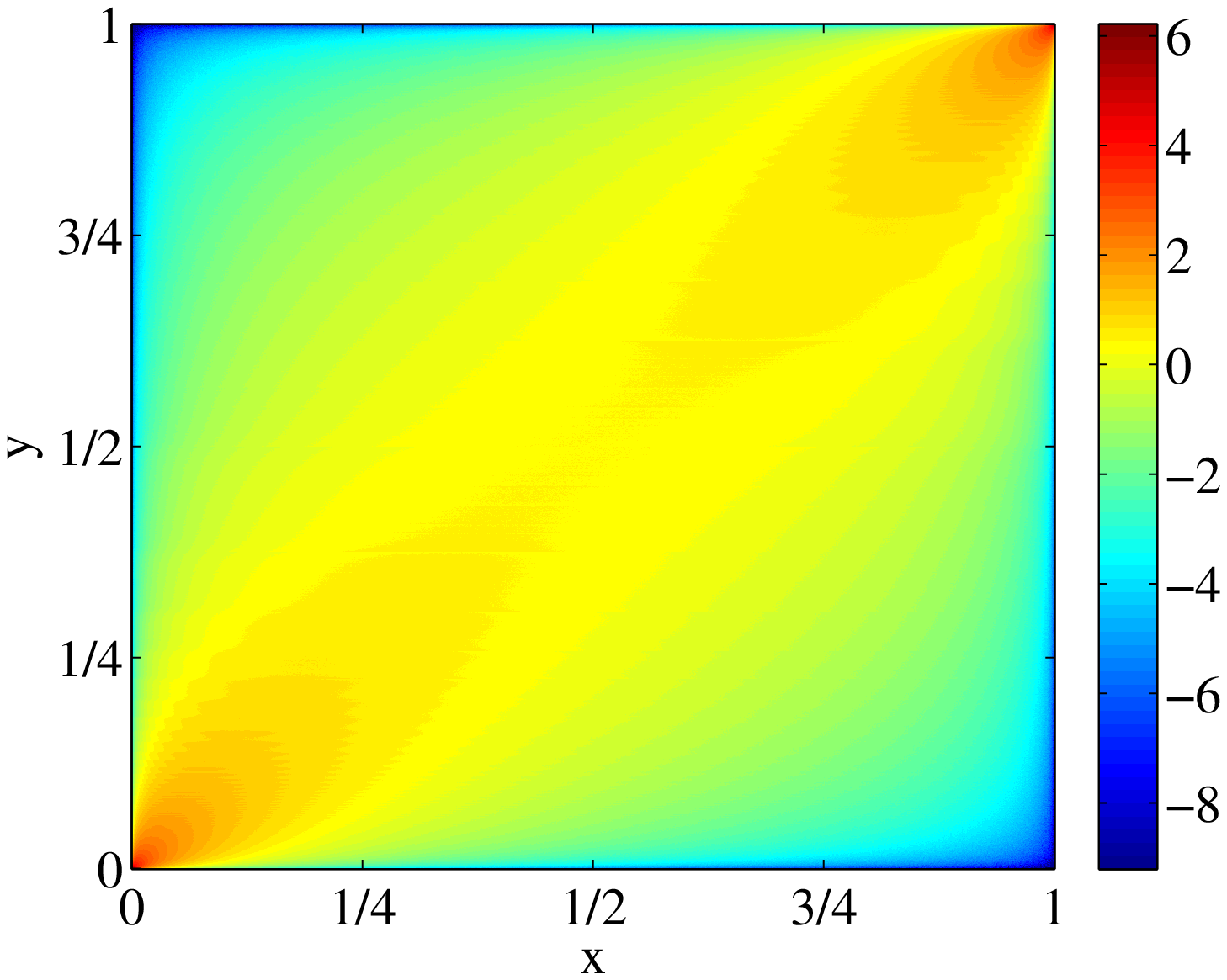}
  }
  \caption{Numerical computations of the invariant density
    $\rho_a(x,y)$, where $a$ takes the values $a = 0.1$, $0.5$, $0.9$, and
    $1$, from left to right and top to bottom. This histogram is computed
    from time series of many trajectories with initial conditions uniformly
   distributed over the square. The grid size is $300\times300$ cells,
   except for the last one which uses a grid of $1000\times1000$
   cells. The colour code uses a logarithmic scale, different for each
   plot, \emph{i.~e.} the legends refer to the natural logarithms of the
   density values. The lower left and upper right corners correspond to
   higher densities, the lower right and upper left corners to lower
   densities.} 
  \label{fig.portraits}
\end{figure}

A remarkable property of $\rho_a$, that can be seen in figure
\ref{fig.portraits}, is that it develops a singularity at the origin as
$a\to1$.

In order to prove this assertion, we consider the partial cumulative
function 
\begin{equation}
r_a(x, y) \equiv \int_0^y\dy \rho_a(x, y).
\label{defcumrho2d}
\end{equation}
From equation (\ref{rho2d}), we obtain the following functional equation
for this quantity~:
\begin{equation}
\fl r_a(x, y) =
\left\{
\begin{array}{l}
  \left(\frac{1 + a}{2} - a x\right)
  r_a\left(\frac{1 + a}{2}x - \frac{a}{2}x^2, 
    \frac{a + 1 - \sqrt{(1 + a)^2 - 8 a y}}{2a}\right),\\
  \hfill 0\leq y \leq 1/2,\\  
  \frac{1 + a}{2} - a x + 
  \left(\frac{1 - a}{2} + a x\right)
  r_a\left(\frac{1}{2} + \frac{1 - a}{2}x + \frac{a}{2}x^2, 
    \frac{\sqrt{(1 + a)^2 - 8 a (1 - y)} - 1 + a}{2a}\right),\\
  \hfill 1/2\leq y \leq 1.  
\end{array} 
\right.\nonumber\\
\label{cumrho2d}
\end{equation}

By construction, we have the properties
\begin{eqnarray}
r_a(x,1) = 1,\label{prop2}\\
\int_0^1\dx r_a(x, y) = y.\label{prop3}
\end{eqnarray}

\begin{figure}[htb]
  \centerline{
    \includegraphics[width=.4\textwidth]{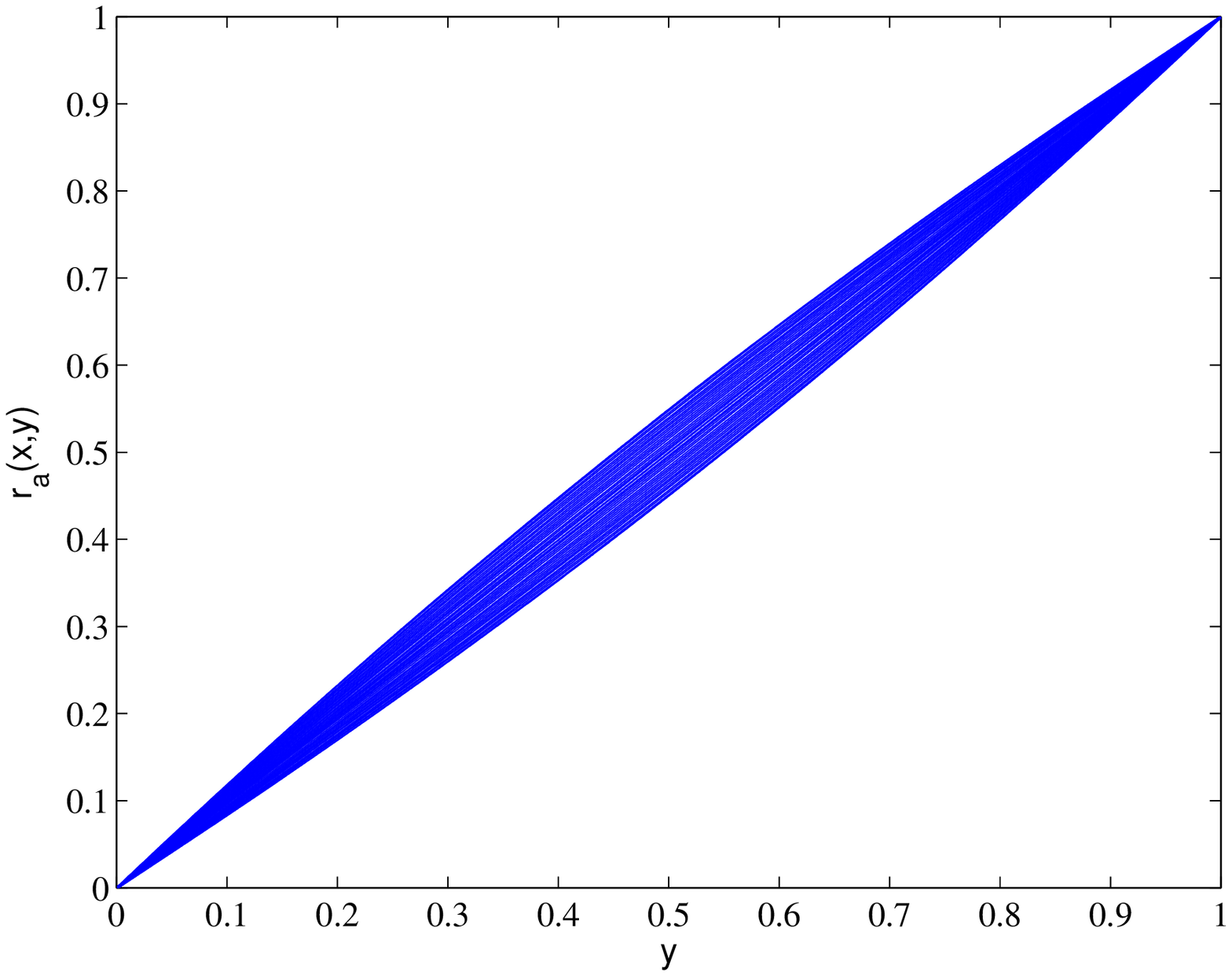}
    \hspace{.5cm}
    \includegraphics[width=.4\textwidth]{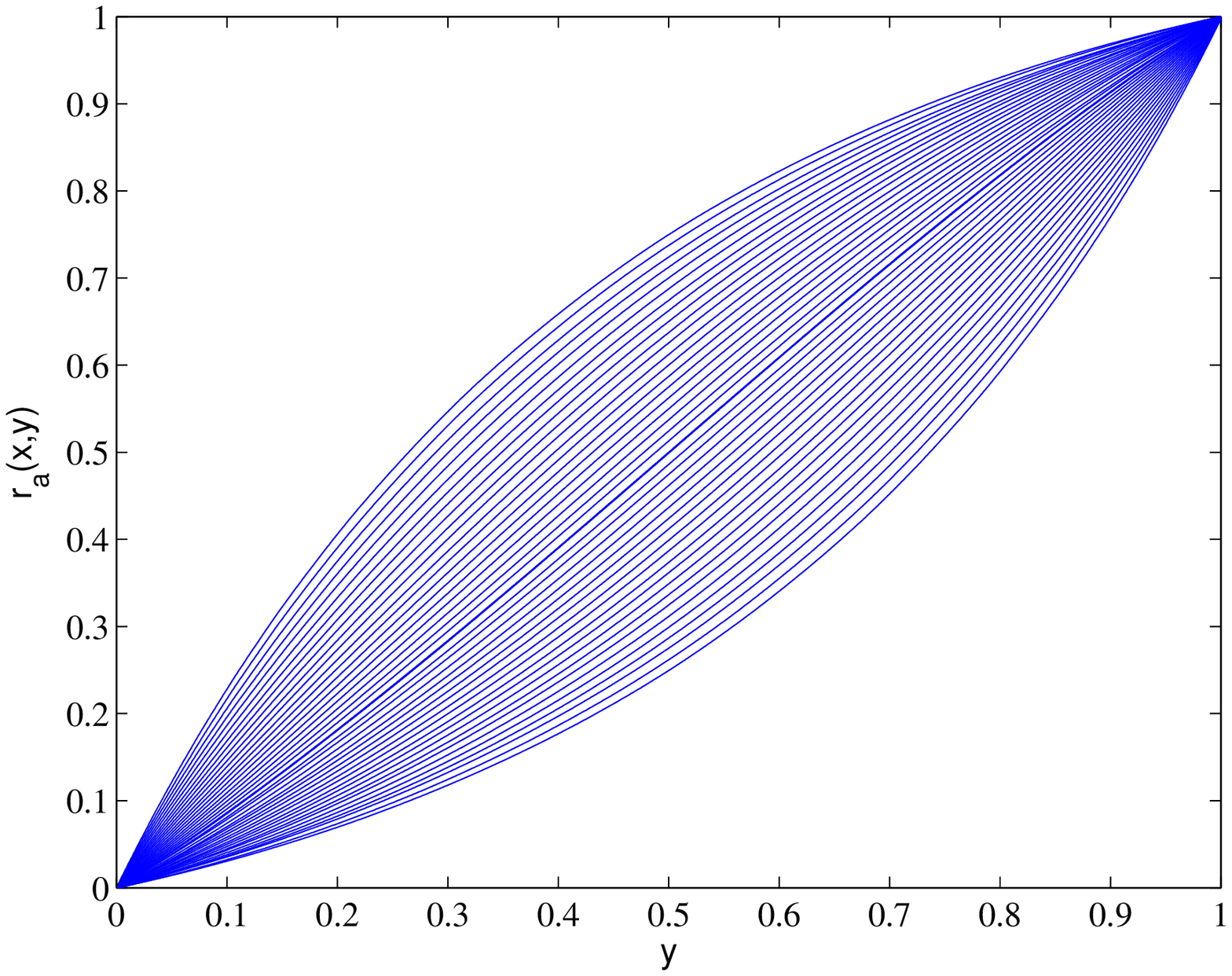}
  }
  \centerline{
    \includegraphics[width=.4\textwidth]{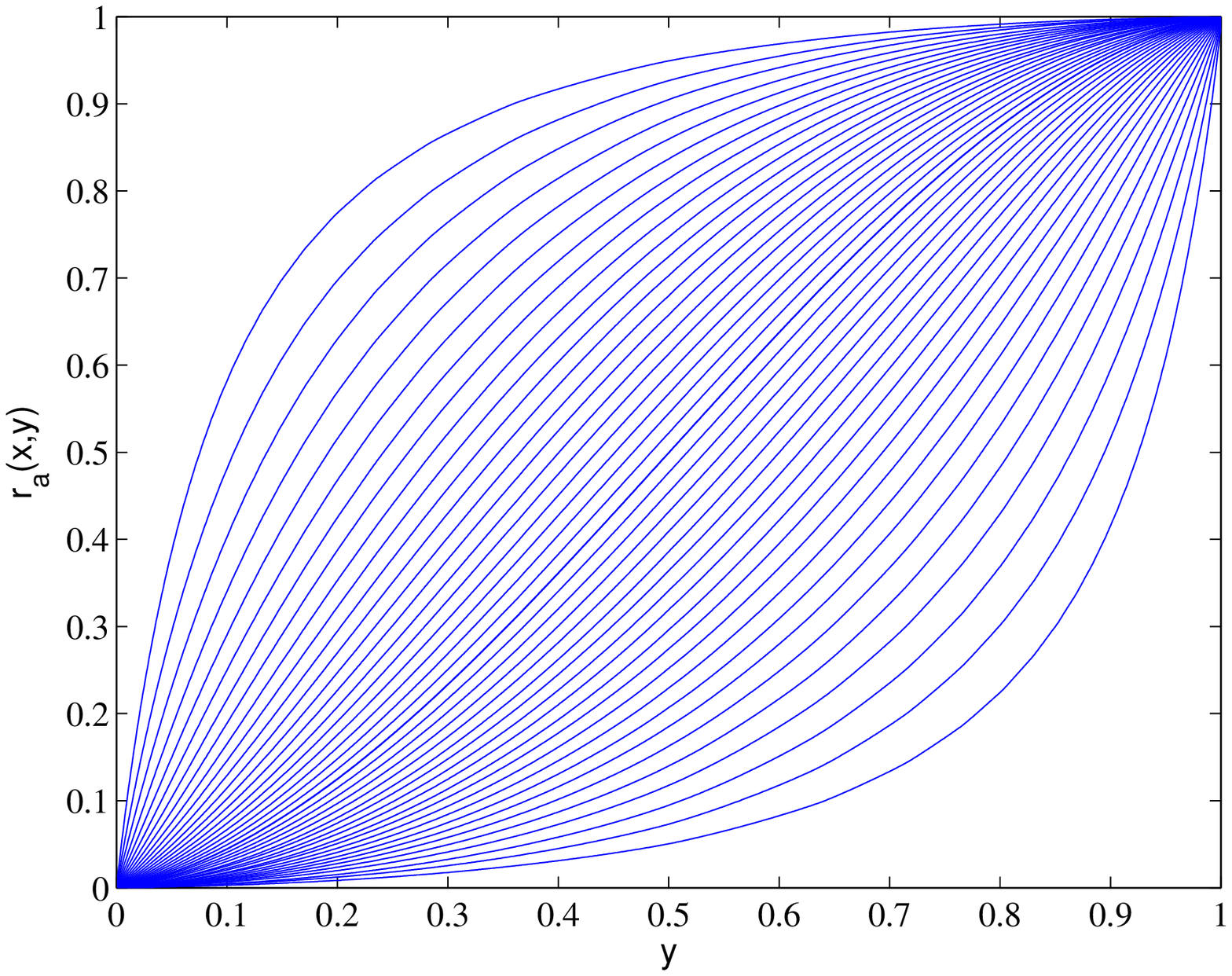}
    \hspace{.5cm}
    \includegraphics[width=.4\textwidth]{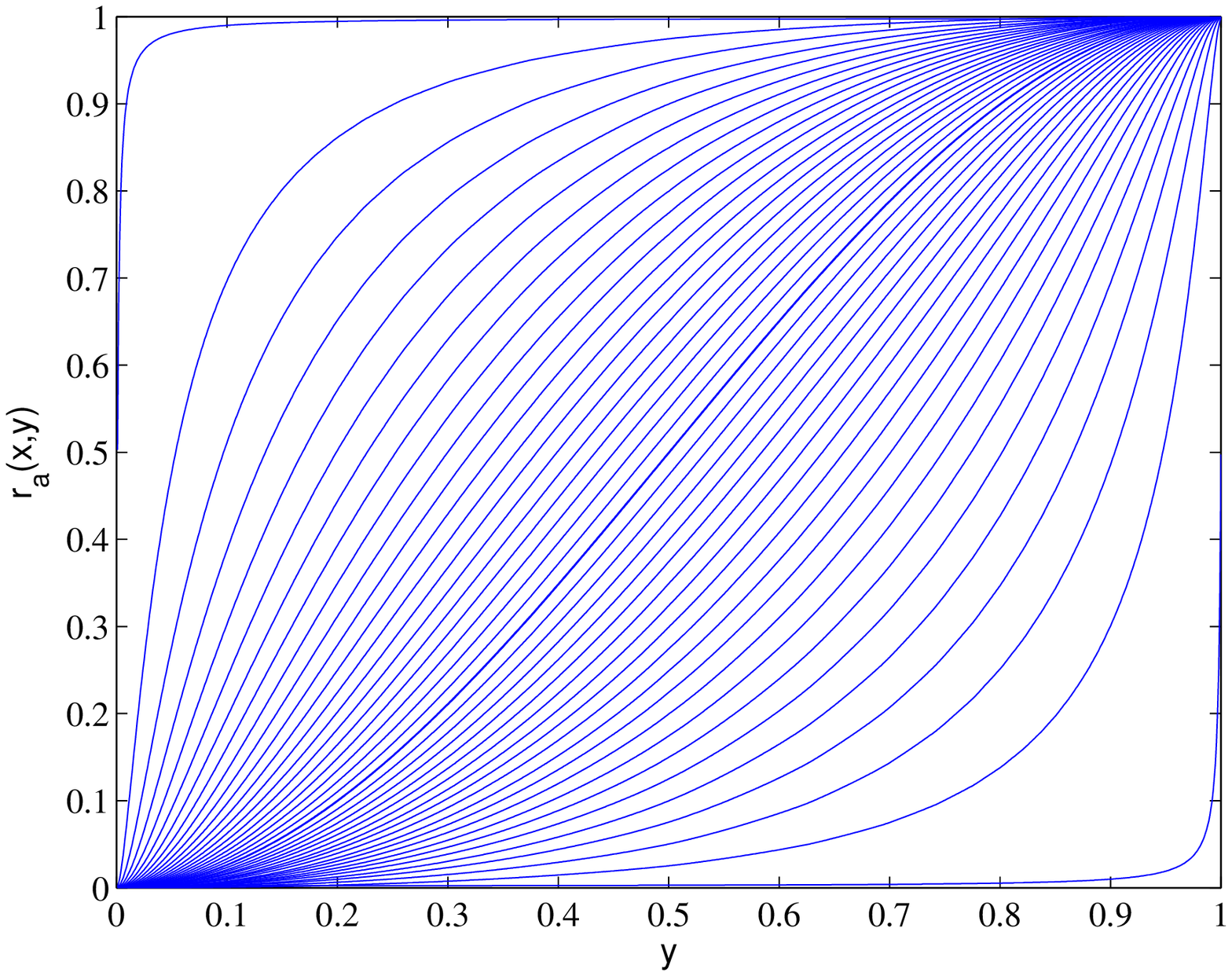}
  }
  \caption{Cumulative functions $r_a(x, y)$ 
    vs. $y$
    , with $a = 0.1$, $0.5$,
    $0.9$ and $1$, from top to bottom and left to right. The different
    curves correspond to different values of $x$ uniformly spread over the
    interval, excluding $x=0$ and $x=1$. The curves are ordered with
    growing values of $x$ from right to left.}
  \label{fig.cumfun}
\end{figure}

Numerical computations of these quantities are displayed in figure
\ref{fig.cumfun}.

The functional Equation (\ref{cumrho2d}) has a form suited to the
computation of the invariant state. Indeed, using equation 
(\ref{prop2}), we can infer the expressions of $r_a(x,y)$ at the points $y$
which are pre-images of $y=1$ with respect to $f_\omega^{(a)}$,
$\omega=0,1$. In particular, it is 
immediate to see that $r_a(x,1/2) = \gal'(x) = (1+a)/2 - ax$.  In general,
the pre-images of $y=1$ are the points $\gaos{n}(1) \equiv \gao{n} \circ
\dots \circ \gao{1}(1)$, where $\uo{n}$ is a compact notation for the
sequence of $n$ digits $\omega_i$, $\omega_i = 0$ or $1$ according to which
branch $\gal$ or $\gar$ is to be iterated.

\begin{lemma}
  Fix $0 < a < 1$. Given any $y_0$ in the unit interval and $\varepsilon
  \ll 1$, one can find an integer $n$ and a sequence $\uo{n}$ such that 
  $\gaos{n}(1) \in (y_0 - \varepsilon, y_0 + \varepsilon)$.
\end{lemma}

The proof is a consequence of the existence of Markov partitions for
the maps (\ref{1dcusp}), $a<1$, and relies on the contraction of the
inverses, $\gal'(x) < 1$ and $\gar'(x) < 1$.
\qed

This implies that $r_a(x,y)$, for fixed $x$ and $a<1$, is completely
determined by equation (\ref{cumrho2d}).

We are specifically interested in those pre-images of $y=1$, which lay
closest to the origin, {\em i.~e.} $y = \gal\circ\dots\circ\gal(1)$.
Thus considering the left branch of equation (\ref{cumrho2d}), we have
\begin{eqnarray}
  r_a\bigg(x, \gazs{n}(1)\bigg) &=& {\gal}'(x)
  r_a\bigg(\gal(x), \gazs{n-1}(1)\bigg),\nonumber\\
  &=& \prod_{k=0}^{n-1} {\gal}'\bigg(\gazs{k}(x)\bigg),
  \nonumber\\
  &=& r_a\bigg(x, \gazs{n-1}(1)\bigg) {\gal}'\bigg(\gazs{n-1}(x)\bigg),
  \nonumber\\
  &=& r_a\bigg(x, \gazs{n-1}(1)\bigg) \left(\frac{a+1}{2} - a 
    \gazs{n-1}(x)\right).
  \label{cumrhorecur}
\end{eqnarray}
These steps easily generalize to any symbolic sequence $\uo{n}$ for which
we can write
\begin{equation}
  r_a\bigg(x, \gaos{n}(1)\bigg) = \gao{n}'(x) r_a\bigg(\gao{n}(x), 
  \gaos{n-1}(1)\bigg) + \omega_n{g^{(a)}_{1-\omega_n}}'(x).
  \label{cumrhorecur2}
\end{equation}
Thus, fixing $x$ and starting at $y=1$, we can use the above equations to
compute points on the curves displayed on the right panels of figure
\ref{fig.cumfun}. 

Setting $x=0$ in equation (\ref{cumrhorecur2}), we
have
\begin{equation}
  r_a\bigg(0, \gaos{n}(1)\bigg) = 
  \gao{n}'(0) r_a\bigg(\gao{n}(0), \gaos{n-1}(1)\bigg) 
  + \omega_n{g^{(a)}_{1-\omega_n}}'(0).
\end{equation}
Thus
\begin{equation}
  r_a\bigg(0, \gaos{n}(1)\bigg) = 
  \left\{
    \begin{array}{l@{\quad}l}
      \frac{a+1}{2} r_a\bigg(0, \gaos{n-1}(1)\bigg),& \omega_n = 0,\\
      \frac{a-1}{2} r_a\bigg(\frac{1}{2}, \gaos{n-1}(1)\bigg) 
      + \frac{a+1}{2},& \omega_n = 1.
    \end{array}
  \right.
\end{equation}
Taking the limit as $a\to1$, we have
\begin{equation}
  r_1\bigg(0, g_{\underline{\omega}_n}^{(1)}(1)\bigg) = 
  \left\{
    \begin{array}{l@{\quad}l}
      r_1\bigg(0, g_{\underline{\omega}_{n-1}}^{(1)}(1)\bigg),
      &\omega_n = 0,\\
      1,& \omega_n = 1.
    \end{array}
  \right.
\end{equation}
Both these two alternatives yield
\begin{equation}
  r_1\bigg(0,g_{\underline{\omega}_n}^{(1)}(1)\bigg) = 1. 
\end{equation}
Indeed, if one amongst the $\omega_{n-1},\dots,\omega_1$ is equal to 1, the
first alternative eventually reduces to the second; if on the other hand 
$\uo{n} = \underline{0}_n$, we get 
\begin{equation}
  r_1\bigg(0,g_{\underline{0}_n}^{(1)}(1)\bigg)
  = r_1\bigg(0,g_0^{(1)}(1)\bigg) = 1.
\end{equation}
Thus
\begin{equation}
  \lim_{a\to1} r_a(0,y) = 1,
  \label{r1x0}
\end{equation}
and, in particular $r_1(0,0) = 1$. Therefore, in the intermittent 
regime, $a=1$, the invariant density has a singularity at the origin,
\begin{equation}
r_1(x,0) = 
\left\{
\begin{array}{l@{\quad}l}
1,&x=0,\\
0,&x>0,
\end{array}
\right.
\label{singrx0}
\end{equation}
which is otherwise absent in the hyperbolic regime, $a<1$, for which
we have $r_a(x,0) = 0$, $x=0$ included. 

The latter property can easily be
checked using simple arguments. Indeed the density at the origin is 
\begin{eqnarray}
\rho_a(0,0) &=& \lim_{n\to\infty} \frac{r_a\bigg(0,\gazs{n+1}(1)\bigg) -
  r_a\bigg(0,\gazs{n}(1)\bigg)}{\gazs{n+1}(1) - \gazs{n}(1)},\nonumber\\
&=& \lim_{n\to\infty} \frac{(1-a)r_a\bigg(0,\gazs{n}(1)\bigg)}
{(1-a)\gazs{n}(1) + a \bigg[\gazs{n}(1)\bigg]^2},\nonumber\\
&=& \lim_{n\to\infty} \frac{r_a\bigg(0,\gazs{n}(1)\bigg)}
{\gazs{n}(1)},\nonumber\\
&=& \lim_{n\to\infty} \frac{[(1+a)/2]^n}
{\gazs{n}(1)}.
\end{eqnarray}
which, as proven in \cite[Theorem 2.1]{Thron}, exists and is finite for $a<1$.
Furthermore, we expect, though we have no formal proof of this result at
this point, that, as $a\to1$, $\rho_a(0,0)$ diverges as
$1/(1-a)$. 

\section{Conclusions}

In this paper, we considered the smooth invariant statistics of
time-reversal symmetric triangular maps of the unit square built upon
anti-symmetric piecewise expanding maps of the unit interval. 

We showed that maps which are diffeomorphically conjugated to piecewise
linear maps have an equilibrium state with the product form, simply
expressed as the product of the derivative of the conjugating map,
evaluated at the two variables.

Maps whose invariant state has the  product form are therefore
exceptional. For piecewise expanding maps that are not diffeomorphically
conjugated to piecewise linear maps, a thorough study of their statistical
properties can only be properly accomplished provided one considers
the map from the interval to the square thus recovering a time-reversal
symmetric map. 

The example of the class of anti-symmetric cusp maps considered in this
paper is revealing in that respect. Though the natural invariant measures of
the one-dimensional maps of this class have uniform densities, even in the
intermittent regime, the equilibrium state of the associated
time-reversible two-dimensional map displays a singularity at the
intermittent fixed point. 

In the non-equilibrium physics literature, time-reversible systems submitted
to non-holonomic constraints have been considered in the context of
non-equilibrium molecular dynamics. Methods were developed over the
last decades using iso-kinetic thermostats under Gauss's principle
of least constraint \cite{EH83}, or Nos\'e-Hoover thermostats \cite{Hoo}. The 
interesting point with regards to the results presented in this paper is
that the equilibrium states of systems subjected to such non-holonomic
constraints are not uniform. This happens because phase-space volumes are
not preserved pointwise, though, in average, they are \cite{GD06}. These
equilibrium states therefore share the properties of the invariant measures
of the maps considered in this paper. Our results suggest that these states
will in general not be factorisable and display a rich structure.

\ack
This research is financially supported by the Belgian Federal Government
(IAP project ``NOSY") and the ``Communaut\'e fran\c caise de Belgique''
(contract ``Actions de Recherche Concert\'ees'' No. 04/09-312). TG is
financially supported by the Fonds de la Recherche Scientifique
F.R.S.-FNRS. 
VB's work is supported in part by the Prodex programme of the European
Space Agency under Contract No. C90241

\end{document}